\documentclass[conference]{IEEEtran}
\IEEEoverridecommandlockouts
\usepackage{cite}
\usepackage{amsmath,amssymb,amsfonts}
\usepackage{algorithmic}
\usepackage{graphicx}
\usepackage{textcomp}
\usepackage[table]{xcolor} 
\usepackage{url}
\usepackage{tablefootnote}
\usepackage{multirow}
\usepackage{float}
\def\BibTeX{{\rm B\kern-.05em{\sc i\kern-.025em b}\kern-.08em
    T\kern-.1667em\lower.7ex\hbox{E}\kern-.125emX}}

\makeatletter
\newcommand*\titleheader[1]{\gdef\@titleheader{#1}}
\AtBeginDocument{%
  \let\st@red@title\@title
  \def\@title{%
    \bgroup\normalfont\large\centering\@titleheader\par\egroup
    \vskip1.5em\st@red@title}
}
\makeatother

\title{A Strictly Bounded Deep Network for Unpaired \\ Cyclic Translation of Medical Images}
\titleheader{A version of this paper is published in 2023 IEEE Statistical Signal Processing Workshop (SSP).
\\ DOI: \url{10.1109/SSP53291.2023.10207960}}
\author{\IEEEauthorblockN{Swati Rai$^1$, Jignesh S. Bhatt$^1$, Sarat Kumar Patra$^2$}
\IEEEauthorblockN{$^1$Indian Institute of Information Technology Vadodara, India \\ $^2$National Institute of Technology Agartala, India}}

\begin{document}

\maketitle
\begin{abstract}
Medical image translation is an ill-posed problem. Unlike existing paired unbounded unidirectional translation networks, in this paper, we consider unpaired medical images and provide a strictly bounded network that yields a stable bidirectional translation. We propose a patch-level concatenated cyclic conditional generative adversarial network (pCCGAN) embedded with adaptive dictionary learning. It consists of two cyclically connected CGANs of 47 layers each; where both generators (each of 32 layers) are conditioned with concatenation of alternate unpaired patches from input and target modality images (not ground truth) of the same organ. The key idea is to exploit cross-neighborhood contextual feature information that bounds the translation space and boosts generalization. The generators are further equipped with adaptive dictionaries learned from the contextual patches to reduce possible degradation. Discriminators are 15-layer deep networks that employ minimax function to validate the translated imagery. A combined loss function is formulated with adversarial, non-adversarial, forward-backward cyclic, and identity losses that further minimize the variance of the proposed learning machine. Qualitative, quantitative, and ablation analysis show superior results on real CT and MRI.
\end{abstract}
\section{Introduction}
Advances in AI and deep learning have fostered developments in statistical learning theory \cite{vapnik2019complete}. It has shown promising successes in medical image denoising \cite{rai2021augmented}, registration \cite{deshpande2019bayesian}, reconstruction \cite{khalid2014single, naderahmadian2018generalized, rai2022accessible}, super-resolution \cite{cherukuri2019deep, rai2022accessible}, and segmentation \cite{pratondo2015robust, rai2022accessible}. There are real-life scenarios where cross-modal information is needed for diagnostic and prognostic purposes, however, a required scanning machine may be inaccessible and/or unaffordable. To this end, a few attempts have been made for medical image translation using paired cross-modal images \cite{armanious2019unsupervised, zhu2020cross, liu2021ct}. Generative adversarial networks (GANs) \cite{NIPS2014_5ca3e9b1} is one of the most popular statistical learning machines that work on the transformation of random fields for cross-domain feature translation in general signal/image processing applications \cite{xu2021domain}. In contrast, cross-domain medical image translation is actually \textit{information translation} task where mapping of contextual information within cross-modality is crucial.
\par In literature, GANs have been used to translate between paired CT and PET \cite{zhu2020cross}, paired non-contrast CT and contrast-enhanced CT \cite{armanious2019unsupervised}, and to synthesize CT from paired MRI \cite{liu2021ct}. However, they need (sufficient) paired images, which are scarcely available in practice. Further, the generators, say in \cite{zhu2020cross, liu2021ct}, are unconditioned, and thus image translation space becomes unbounded leading to inconsistent generalization. 
\par In this paper, to address the ill-posedness of unpaired CT-MRI translation (of the same organ), we propose a strictly bounded deep cyclic CGAN. It generates characteristic information of target modality while preserving the characteristics of input modality when back translated. Note that the term unpaired refers to training without the need for corresponding ground truth target image. The translation is achieved by conditioning (bounding) the generator networks with concatenated alternate patches of unpaired CT/MRI imagery to restrict the solution space while embedding with dictionary learning (DiL) logic to perform the simultaneous restoration. A combined loss function, incorporating adversarial, non-adversarial, forward-backward cyclic, and identity losses, is proposed to further ensure stability of neural network. We aim to construct a prototype neural network for simulating a target image modality from the given unpaired modality.
\section{Proposed Approach}
\begin{figure}
    \centering
    \includegraphics[width=\linewidth]{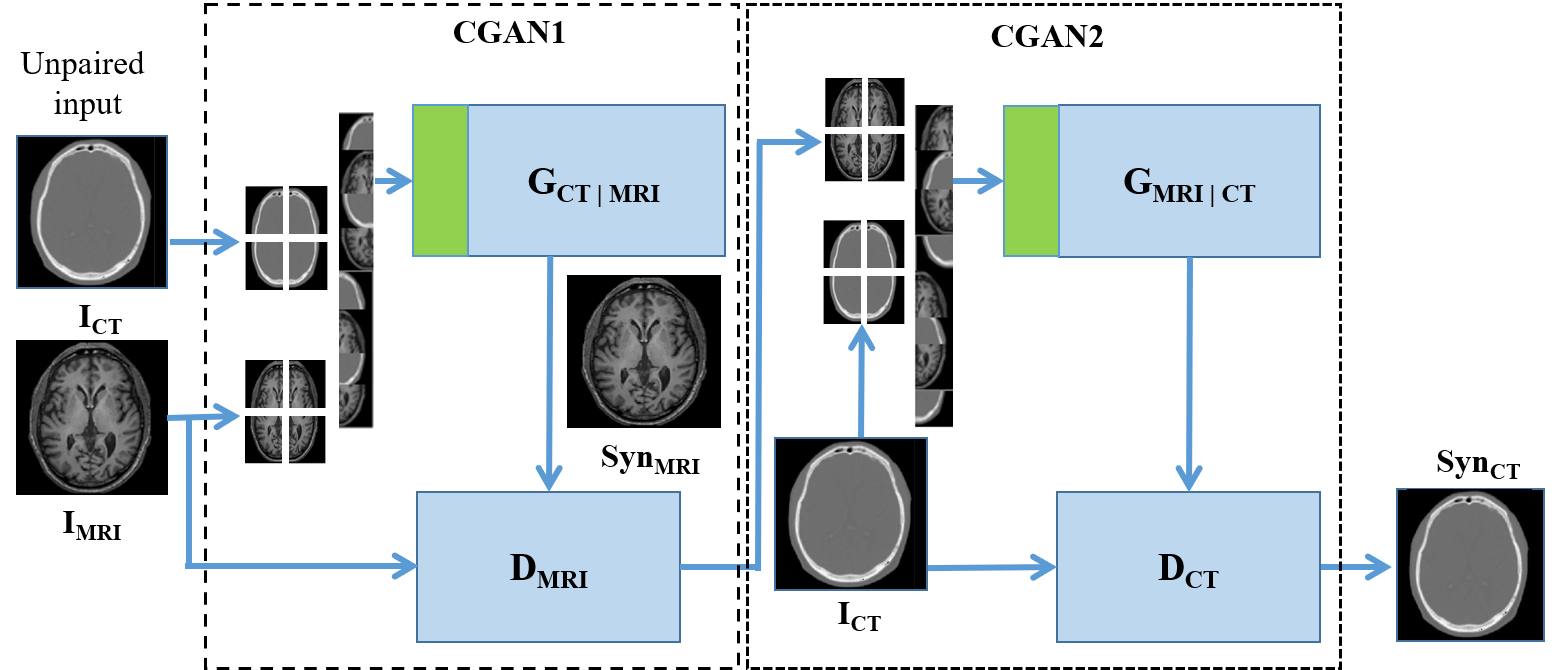}
    \caption{Schematic of proposed approach for CT-MRI-CT translation.}
    \label{3.1}
\end{figure}

This section describes patch-level cyclic conditional generative adversarial network (pCCGAN) embedded with DiL based restoration constructed for unpaired cyclic CT-MRI translation. Given an unpaired input (CT-MRI), our network constructs corresponding target modality (MRI-CT). We assume that input and target images are of the same organ. As shown in Fig. \ref{3.1}, our network consists of two CGANs connected conditionally cyclic. Here, CGAN1 is trained by conditioning on unpaired alternate patches of input-target, while CGAN2 is trained on alternate patches of translated target-input. 
\begin{figure*}
    \centering
    \includegraphics[width=\linewidth]{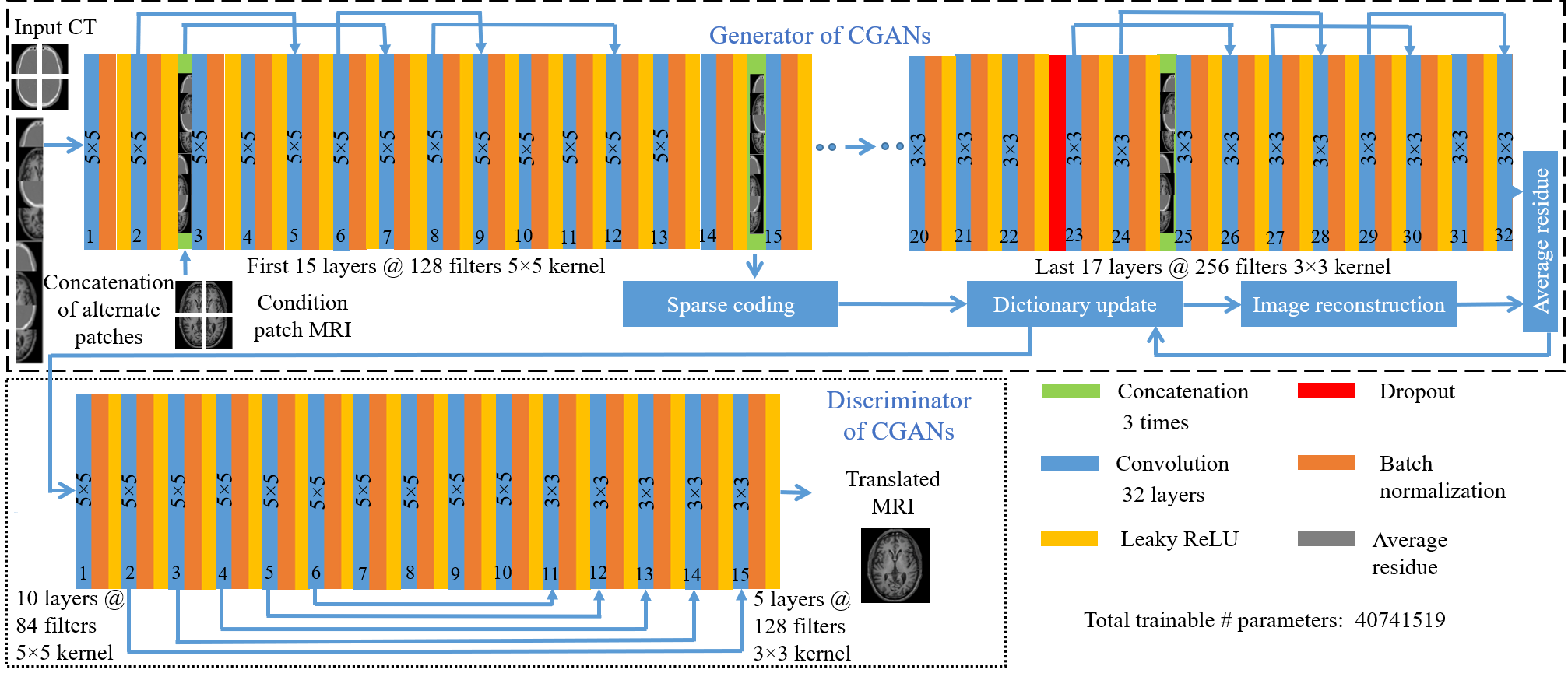}
    \caption{Proposed deep neural architecture of CGANs in the schematic (Fig. \ref{3.1}).}
    \label{3.2}
\end{figure*}
\par Referring to Fig. \ref{3.1} for CT-MRI-CT translation, the network considers input CT image $I_{CT}$ and translate it to an equivalent MRI image $Syn_{MRI}$, and then back translate to $Syn_{CT}$. In CGAN1, concatenated alternate patches of $I_{CT}$ and MRI image $I_{MRI}$ (unpaired) are presented at the generator $G_{CT\:|\:MRI}$. For a 
scenario of CT to MRI translation where all images are of size $n\times n$ pixels, with patches of different sizes, the concatenation of alternate patches given to generator $G_{CT\:|\:MRI}$ is as follows:
\begin{equation}
\begin{split}
patch^1_{CT}\|patch^1_{MRI}\|patch^2_{CT}\|patch^2_{MRI}\|\cdot\cdot\cdot \\ \|patch^{n}_{CT}\|patch^{n}_{MRI} \rightarrow G_{CT\:|\:MRI}. 
\end{split}
\end{equation}
See that, it provides patch-level conditioning by target modality in order to effectively leverage on learning the cross-contextual information features. This key idea better helps bounding the generator (solution) space and helps performing stable cyclic translation of medical imagery (random fields). Inspired from our work \cite{rai2021augmented}, the adaptive dictionaries are trained on the characteristics of target modality, and further constrain in restoration of $Syn_{MRI}$ in $G_{CT\:|\:MRI}$. It is then presented to the discriminator $D_{MRI}$ already trained to validate MRI images. In turn, the CGAN2 performs back translation from $Syn_{MRI}$ to $Syn_{CT}$ in order to establish equivalence with respect to the $I_{CT}$ at input of CGAN1. We run through both the CGANs during training and validation, and for test translation, a single CGAN is suffice.
\par The formulated loss function comprises following terms,
\begin{equation}
    L_{Total} = \gamma \: (L_{CGAN1} + L_{CGAN2}) + L_{cyc} + L_{id},
\end{equation}
where $L_{Total}$ is the resultant loss value, $\gamma$ is an adjustable parameter according to target modality, $L_{CGAN1}$ and $L_{CGAN2}$ are the adversarial losses for two CGANs and defined as:
\begin{equation}
\begin{split}
    L_{CGAN1}(G_{CT|MRI}, D_{MRI}, I_{CT}, I_{MRI}) =  \\ \mathbb{E}_{I_{MRI}\sim p_{data}(MRI)} [D_{MRI}(I_{MRI})_{2}^{2}] + \\ \mathbb{E}_{I_{CT}\sim p_{data}(CT)} [1 - D_{MRI}(G_{CT|MRI}(I_{CT}))_{2}^{2}],
\end{split}
\end{equation}
\begin{equation}
\begin{split}
    L_{CGAN2}(G_{MRI|CT}, D_{CT}, Syn_{MRI}, I_{CT}) =  \\ \mathbb{E}_{I_{CT}\sim p_{data}(CT)} [D_{CT}(I_{CT})_{2}^{2}] + \mathbb{E}_{Syn_{MRI}\sim p_{data}(MRI)} \\ [1 - D_{CT}(G_{MRI|CT}(Syn_{MRI}))_{2}^{2}],
\end{split}
\end{equation}
where $\mathbb{E}(\cdot)$ indicates conditional expectation.
The network learns local information by minimizing non-adversarial losses, i.e., cyclic loss $L_{cyc}$, and identity loss $L_{id}$. Here, $L_{cyc}$ comprises forward/backward cyclic losses and defined as:
\begin{equation}
    \begin{split}
        L_{cyc}(G_{CT | MRI}, G_{MRI | CT}) = \mathbb{E}_{I_{CT}\sim p_{data}(CT)}(G_{MRI | CT}\\(G_{CT | MRI}(I_{CT})) - I_{CT}) + \mathbb{E}_{I_{MRI}\sim p_{data}(MRI)}(G_{CT|MRI}\\(G_{MRI|CT}(I_{MRI})) - I_{MRI}). 
    \end{split}
\end{equation}
The $L_{id}$ is defined by presenting the input image directly to the $G_{MRI\:|\:CT}$ (CGAN2) conditioned on the CT as target modality, and giving MRI image as input to the $G_{MRI\:|\:CT}$ (CGAN1). It is defined as:
\begin{equation}
    \begin{split}
        L_{id}(G_{CT | MRI}, G_{MRI | CT}) = \mathbb{E}_{I_{MRI}\sim p_{data}(MRI)}(G_{CT | MRI}\\(I_{MRI}) - I_{MRI}) + \mathbb{E}_{I_{CT}\sim p_{data}(CT)}\\(G_{MRI|CT}(I_{CT}) - I_{CT}). 
    \end{split}
\end{equation}
Similarly, for MRI-CT-MRI translation, we suitably modify functional blocks shown in Fig. \ref{3.1} and redefine (1) to (6).
\begin{figure*}
    \centering
    \includegraphics[width=\linewidth]{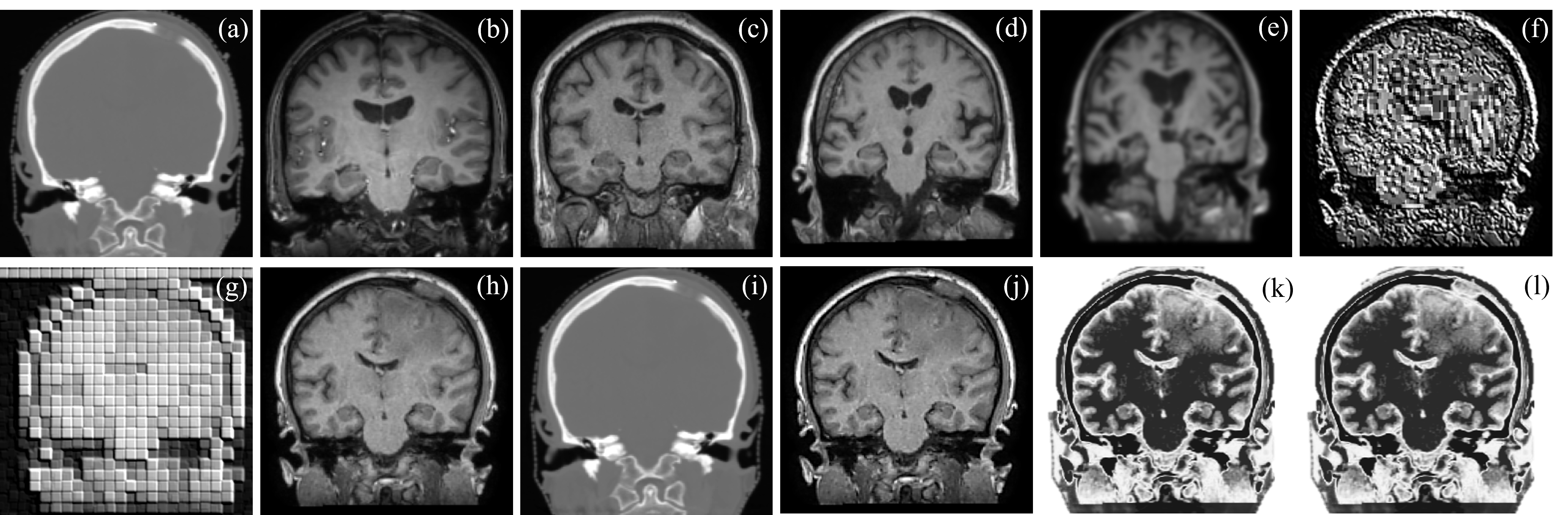}
    \caption{(with ground truth) Qualitative analysis of \textcolor{blue}{unpaired CT-MRI-CT translation of real brain images}: (a) given CT, (b) translated MRI by baseline CGAN, (c) an unpaired MRI for conditioning; translated MRI when conditioned by: (d) random target image, (e) average target image, (f) sample PDF, (g) $8\times8$ patch, (h) \textcolor{blue}{$16\times16$ patch}; (i) \textcolor{blue}{back translated CT from (h)}, (j) ground truth MRI for (a); difference image: (k) between (a)-(j), (l) between (a)-(h).}
    \label{3.3}
\end{figure*}
\begin{table*}
\centering
\caption{Average scores and ablation analysis over 100 random test images of brain and pelvis images.}
\resizebox{0.9\linewidth}{!}{%
\begin{tabular}{|c|cccccc|cccccc|}
\hline
\multirow{3}{*}{\begin{tabular}[c]{@{}c@{}}Proposed approach \\ \textbf{under different conditioning}\end{tabular}} & \multicolumn{6}{c|}{\textbf{CT-MRI-CT}}                                                                                                 & \multicolumn{6}{c|}{\textbf{MRI-CT-MRI}}                                                                                               \\ \cline{2-13} 
& \multicolumn{2}{c|}{\textbf{RMSE$\boldsymbol{\downarrow}$}}   & \multicolumn{2}{c|}{\textbf{PSNR$\boldsymbol{\uparrow}$}}   & \multicolumn{2}{c|}{\textbf{SSIM$\boldsymbol{\uparrow}$}}                      & \multicolumn{2}{c|}{\textbf{RMSE$\boldsymbol{\downarrow}$}}   & \multicolumn{2}{c|}{\textbf{PSNR$\boldsymbol{\uparrow}$}}   & \multicolumn{2}{c|}{\textbf{SSIM$\boldsymbol{\uparrow}$}}                      
 \\ \cline{2-13} 
& \multicolumn{1}{c|}{\begin{tabular}[c]{@{}c@{}}Without \\ DiL\end{tabular}} & \multicolumn{1}{c|}{\begin{tabular}[c]{@{}c@{}}With \\ DiL\end{tabular}} & \multicolumn{1}{c|}{\begin{tabular}[c]{@{}c@{}}Without \\ DiL\end{tabular}} & \multicolumn{1}{c|}{\begin{tabular}[c]{@{}c@{}}With \\ DiL\end{tabular}} & \multicolumn{1}{c|}{\begin{tabular}[c]{@{}c@{}}Without \\ DiL\end{tabular}} & \begin{tabular}[c]{@{}c@{}}With \\ DiL\end{tabular} & \multicolumn{1}{c|}{\begin{tabular}[c]{@{}c@{}}Without \\ DiL\end{tabular}} & \multicolumn{1}{c|}{\begin{tabular}[c]{@{}c@{}}With \\ DiL\end{tabular}} & \multicolumn{1}{c|}{\begin{tabular}[c]{@{}c@{}}Without \\ DiL\end{tabular}} & \multicolumn{1}{c|}{\begin{tabular}[c]{@{}c@{}}With \\ DiL\end{tabular}} & \multicolumn{1}{c|}{\begin{tabular}[c]{@{}c@{}}Without \\ DiL\end{tabular}} & \begin{tabular}[c]{@{}c@{}}With\\  DiL\end{tabular} \\ \hline
Unconditional CGAN inspired from \cite{armanious2019unsupervised, liu2021ct}     & \multicolumn{1}{c|}{17.732}    & \multicolumn{1}{c|}{17.732}    & \multicolumn{1}{c|}{38.473}                   & \multicolumn{1}{c|}{38.473}   & \multicolumn{1}{c|}{0.772}    & 0.772         & \multicolumn{1}{c|}{16.992}                          & \multicolumn{1}{c|}{16.992}   & \multicolumn{1}{c|}{39.473}   & \multicolumn{1}{c|}{39.473}     & \multicolumn{1}{c|}{0.718}          & 0.718                                                                 \\ \hline
Random target image   & \multicolumn{1}{c|}{20.274}     & \multicolumn{1}{c|}{18.241}      & \multicolumn{1}{c|}{28.267}                & \multicolumn{1}{c|}{32.552}   & \multicolumn{1}{c|}{0.495}      & 0.693      & \multicolumn{1}{c|}{21.113}                           & \multicolumn{1}{c|}{18.225}   & \multicolumn{1}{c|}{29.177}     & \multicolumn{1}{c|}{33.482}     & \multicolumn{1}{c|}{0.503}        & 0.691                                                                 \\ \hline
Average image from 100 targets   & \multicolumn{1}{c|}{33.648}     & \multicolumn{1}{c|}{30.991}     & \multicolumn{1}{c|}{25.145}       & \multicolumn{1}{c|}{29.362}   & \multicolumn{1}{c|}{0.466}      & 0.551          & \multicolumn{1}{c|}{31.648}                       & \multicolumn{1}{c|}{29.875}   & \multicolumn{1}{c|}{25.228}      & \multicolumn{1}{c|}{28.429}    & \multicolumn{1}{c|}{0.401}        & 0.502                                                                 \\ \hline
Sample PDF from 100 targets      & \multicolumn{1}{c|}{92.986}     & \multicolumn{1}{c|}{91.189}     & \multicolumn{1}{c|}{6.667}        & \multicolumn{1}{c|}{8.618}   & \multicolumn{1}{c|}{0.109}      & 0.117               & \multicolumn{1}{c|}{90.166}                  & \multicolumn{1}{c|}{88.283}   & \multicolumn{1}{c|}{6.291}      & \multicolumn{1}{c|}{8.422}     & \multicolumn{1}{c|}{0.102}        & 0.108                                                                 \\ \hline
Concatenated alternate 8x8 patches   & \multicolumn{1}{c|}{88.653}    & \multicolumn{1}{c|}{85.446}   & \multicolumn{1}{c|}{10.293}    & \multicolumn{1}{c|}{12.139}   & \multicolumn{1}{c|}{0.182}    & 0.219         & \multicolumn{1}{c|}{85.568}                          & \multicolumn{1}{c|}{83.592}   & \multicolumn{1}{c|}{11.001}      & \multicolumn{1}{c|}{13.193}     & \multicolumn{1}{c|}{0.113}       & 0.221                                                                 \\ \hline
\cellcolor{gray!30}{Concatenated alternate 16x16 patches}    & \multicolumn{1}{c|}{\cellcolor{gray!30}{14.741}}      & \multicolumn{1}{c|}{\cellcolor{gray!30}{\textbf{12.112}}}                       & \multicolumn{1}{c|}{\cellcolor{gray!30}{39.451}}   & \multicolumn{1}{c|}{\cellcolor{gray!30}{\textbf{45.624}}}         & \multicolumn{1}{c|}{\cellcolor{gray!30}{0.808}}                    & \textbf{\cellcolor{gray!30}{0.912}}         & \multicolumn{1}{c|}{\cellcolor{gray!30}{15.033}}       & \multicolumn{1}{c|}{\cellcolor{gray!30}{\textbf{11.241}}}   & \multicolumn{1}{c|}{\cellcolor{gray!30}{39.882}}    & \multicolumn{1}{c|}{\cellcolor{gray!30}{\textbf{48.347}}}     & \multicolumn{1}{c|}{\cellcolor{gray!30}{0.812}}         & \textbf{\cellcolor{gray!30}{0.922}}                                                                 \\ \hline
Concatenated alternate 32x32 patches     & \multicolumn{1}{c|}{17.905}     & \multicolumn{1}{c|}{15.667}                                & \multicolumn{1}{c|}{37.775}            & \multicolumn{1}{c|}{40.196}     & \multicolumn{1}{c|}{0.756}                                 & 0.899                                 & \multicolumn{1}{c|}{17.781}     & \multicolumn{1}{c|}{13.739}                                & \multicolumn{1}{c|}{38.269}            & \multicolumn{1}{c|}{43.661}     & \multicolumn{1}{c|}{0.756}                                 & 0.909                                                                 \\ \hline
Concatenated alternate 64x64 patches     & \multicolumn{1}{c|}{43.669}     & \multicolumn{1}{c|}{42.843}                                & \multicolumn{1}{c|}{23.113}            & \multicolumn{1}{c|}{25.781}     & \multicolumn{1}{c|}{0.332}                                 & 0.495                                 & \multicolumn{1}{c|}{39.184}     & \multicolumn{1}{c|}{38.372}                                & \multicolumn{1}{c|}{24.681}            & \multicolumn{1}{c|}{27.441}     & \multicolumn{1}{c|}{0.392}                                 & 0.511                                                                 \\ \hline
\end{tabular}}\par
\smallskip
All the experiments run for 800 epochs in python 3.7 on PARAM Shavak high-performance computing (HPC) supercomputer with 96 GB RAM.
\end{table*}
\subsection{Network and Training}
We propose novel deep neural architecture of CGANs (Fig. \ref{3.2}) used in our approach (Fig. \ref{3.1}). Referring to Fig. \ref{3.2}, generator is a deep convolutional neural network (CNN) equipped with DiL; that simultaneously restores the image. The network consists of two deep CGANs each with 47 layers. A generator in CGAN consists of 32 layers out of which the first 15 convolution layers have 128 filters each of size $5\times5$ pixels while the last 17 layers have 256 filters of size $3\times3$ pixels. The input to these layers are conditioned by concatenating alternate patches of target image at the 3$^{rd}$, 15$^{th}$, and 25$^{th}$ layers to bound the generator space to the target modality space. See that in the first 15 layers, the convolution layers simultaneously learn the intrinsic contextual cross-domain information features with the help of cross-neighborhood conditioning and thus enhance generalization. Then, the target image modality patterns are again concatenated patch-wise on the learned cross-neighborhood information features and presented to the next 17 layers with 256 filters of size $3\times3$ pixels. It further helps the network to influence statistical characteristics of the target modality in the learned information features and helps improve cyclic translation. Except the last layer of the generator, each convolution layer is followed by batch normalization (BN) to enable higher learning rates and minimize the internal covariate shift. It is succeeded by a leaky rectified linear unit (leaky ReLU) to avoid the dying ReLU problems and to speed up the training process. The dropout layer and skip connections are inserted to avoid possible overfitting and vanishing/shattered gradients, respectively. 
\begin{figure*}
    \centering
    \includegraphics[width=0.8\linewidth]{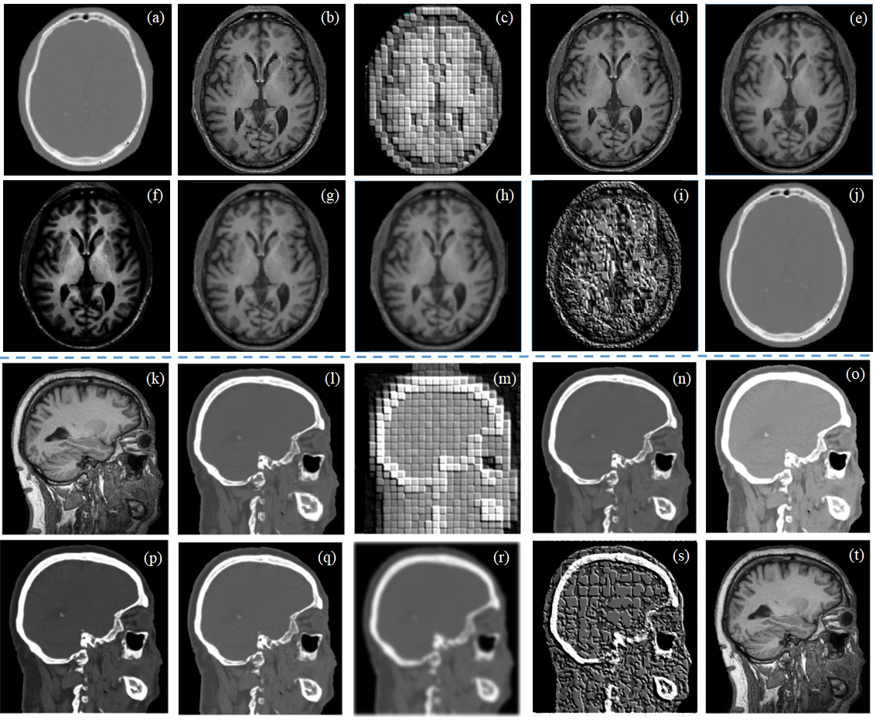}
    \caption{(without ground truth) Qualitative analysis of unpaired image translation of brain CT-MRI-CT ((a) to (j)), and cross-sectional face MRI-CT-MRI ((k) to (t)): \textcolor{blue}{(a-b) Unpaired CT-MRI and (k-l) unpaired MRI-CT}; translated (c) MRI and (m) CT by $8\times8$; \textcolor{blue}{translated (d) MRI and (n) CT by $16\times16$}; translated (e) MRI and (o) CT by $32\times32$; translated (f) MRI and (p) CT by $64\times64$; translated (g) MRI and (q) CT by random target modality image; translated (h) MRI and (r) CT by average image of target modality; translated (i) MRI and (s) CT by sample target PDF; and \textcolor{blue}{back translated (j) CT from (d) and back translated (t) MRI from (n)}.}
    \label{3.4}
\end{figure*}
\par Simultaneously, the patch-level concatenated vector at the 15$^{th}$ layer is given to the sparse coding dictionaries. Here the sparse coefficients are learned from the concatenated patches and initial dictionary. We initialize the dictionary with our work on medical image denoising \cite{rai2021augmented}. With the help of updated sparse coefficients and concatenated patches, the network updates the initial dictionary. Residue is learned from the adaptive dictionaries and convolution layers, in turn, summarize them to unsupervisely learn the degradation present in both high-frequency and low-frequency regions \cite{rai2021unsupervised}. Interestingly, it also avoids over-optimization, which in turn bypasses the mode collapse problem commonly found in GANs. 

\begin{figure*}
    \centering
    \includegraphics[width=\linewidth]{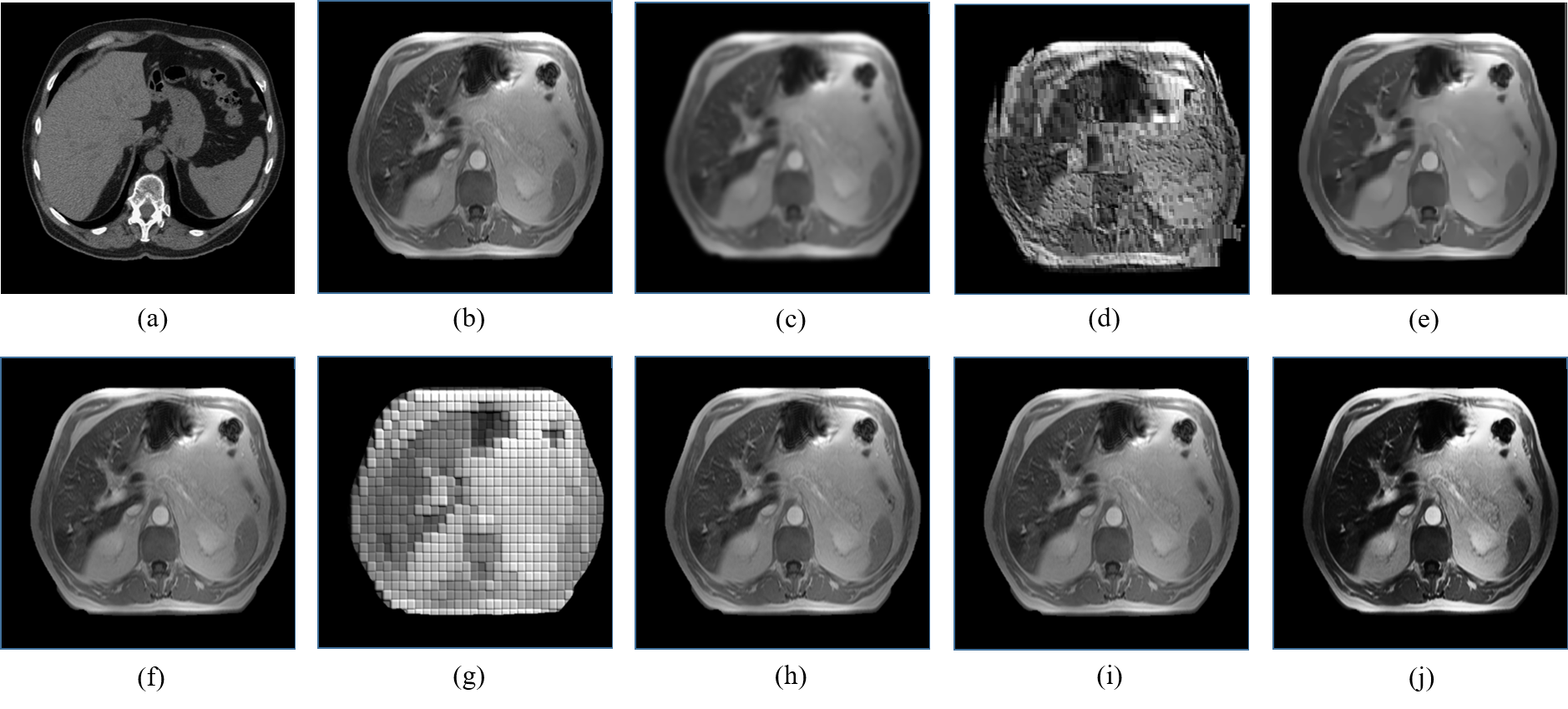}
    \caption{CT-MRI-CT unpaired image translation of a 41-year male abdomen using the proposed pCCGAN under different conditioning scenarios. (a) and (f) are real CT and MRI images, respectively, (b) Conditioning by images, (c) conditioning by average image, (d) conditioning by PDF, (e) CCGAN without augmentation, (g) conditioning by a patch of size $8\times8$ pixels, (h) conditioning by patch of size $16\times16$ pixels, (i) conditioning by patch of size $32\times32$ pixels, and (j) conditioning by patch of size $64\times64$ pixels.}
    \label{3.6}
\end{figure*}
\par The discriminator of our CGANs (Fig. \ref{3.2}) is a 15-layer deep CNN that takes the translated image from the generator and validates it for target modality. It consists of the first 10 convolution layers with 84 filters of size $5\times5$ pixels followed by BN and leaky ReLU to compare the distribution of the modalities. The last 5 convolution layers use 128 filters with a kernel of size $3\times3$ pixels in each layer.
\section{Results}
In this section, we discuss the results obtained by the proposed approach using the publicly available benchmark paired real MRI-CT datasets \cite{thummerer_adrian_2023_7260705} and unpaired datasets \cite{CTMRI}. It includes (train/ validation/ test): Images of \#brain (5000/ 800/ 900); \#pelvis (500/ 100/ 150); and \#face (1080/ 200/ 220). In this experiment, all images are of size $256\times 256$ pixels. We employ Adam optimization with $3\:e^{-4}$ learning rate, and the adjustable parameter $\gamma$ in (2) is empirically set to 1.5. We present qualitative, quantitative, and ablation results with different conditioning scenarios as well as with baseline CGAN inspired from state-of-the-art \cite{armanious2019unsupervised, liu2021ct}. 
\begin{figure*}
    \centering
    \includegraphics[width=0.6\linewidth]{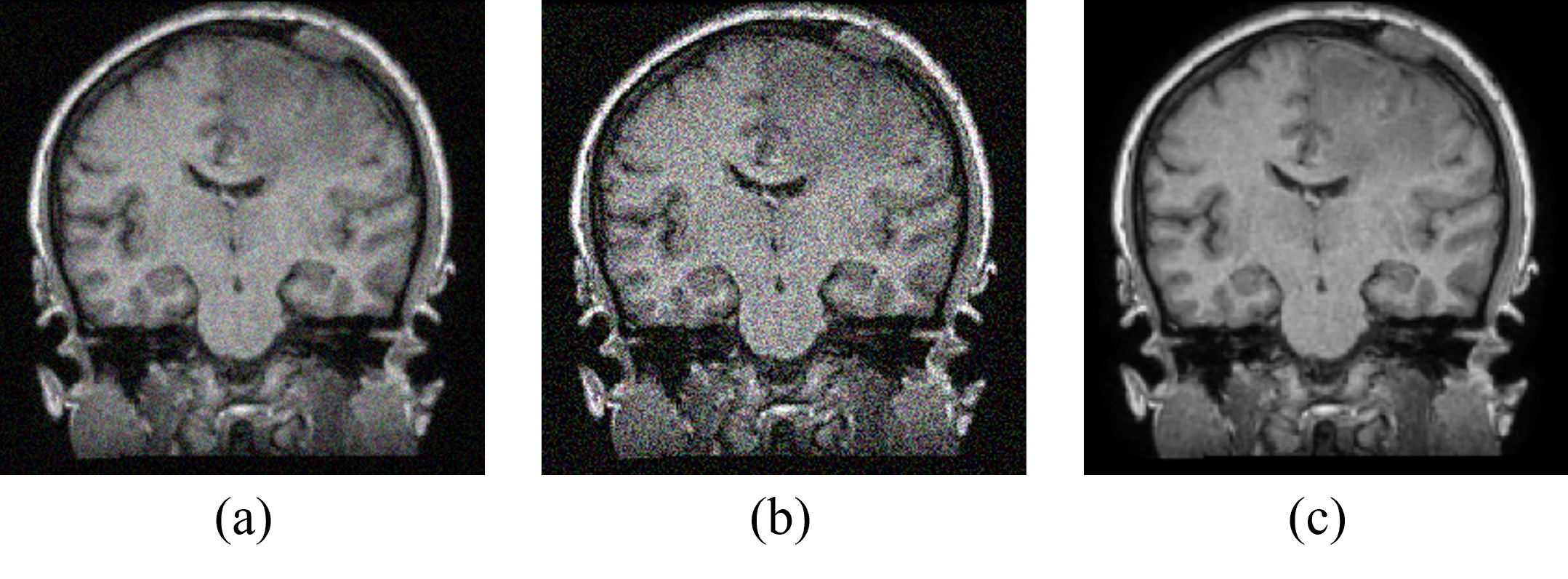}
    \caption{Ablation result for DiL: (a) given MRI, (b) reconstructed MRI without DiL, and (c) reconstructed MRI \textcolor{blue}{with DiL}.}
    \label{3.5}
\end{figure*}
\par We have trained the network by conditioning it in different scenarios: (a) \textit{Patches of target images}: The input and target images are patch-wise concatenated with different patch sizes, i.e., $8\times8$, $16\times16$, $32\times32$, and $64\times64$ pixels in an alternate manner to condition the generator. (b) \textit{Random target image}: We condition the generator with a random image from a set of targets. (c) \textit{Average image from a set of targets}: The generator is conditioned with pixel-by-pixel averaging 100 target images. (d) \textit{Sample probability distribution function (PDF) of target modality}: We condition the generator by a sample PDF constructed with 100 target images.  

\par Fig. \ref{3.3} displays a qualitative result obtained using the pCCGAN by different conditioning scenarios for unpaired cyclic CT-MRI translation. Fig. \ref{3.3} (j) shows ground truth MRI. One may observe that the MRI generated by conditioning of $16\times16$ patches (Fig. \ref{3.3} (h)) optimally utilizes the cross-neighborhood pixel information and achieves superior translation and restoration. Quantitative and ablation test analysis for 100 test image translations are presented in Table I. This confirms the superiority of proposed approach conditioned with $16\times16$ concatenated alternate patches with DiL. Further, Fig. \ref{3.4} (a) to (t) display a few qualitative results for unpaired cyclic CT-MRI-CT and MRI-CT-MRI translation when the ground truth is not present. One may observe from Fig. \ref{3.4} that the proposed approach performs superior unpaired translation even in the absence of ground truth. 

\par An additional CT to MRI abdomen image translation by all the conditioning scenarios is shown in Fig \ref{3.6} (a) to (j).  The real input abdomen CT image and corresponding target abdomen MRI image are shown in Fig. 5 (a) and Fig. 5 (f), respectively. Fig. 5 (b) presents the generated image from the proposed pCCGAN conditioned with images. One can observe that the edges are not clearly defined especially at the bottom part of the generated image. The output image generated by the conditioning of the average MRI image is displayed in Fig. 5 (c). Here, one can see that the image is a blur which leads to missing minute details. Fig. 5 (d) shows the result obtained by giving a PDF of an MRI image as the condition; one can make out from the image that the proposed CCGAN is not able to translate the input CT image to an MRI image. The output generated by the simple pCCGAN without augmentation is displayed in Fig. 5 (e). In this case, the image generated overshadows the low-frequency regions and the high-frequency region is enhanced. Fig. 5 (g) presents the image generated by the proposed pCCGAN when a patch of $8\times8$ pixels is the condition provided. The network tries to translate the CT into MRI, however, the image remains patchy. The image generated by the conditioning of $16\times16$ pixels size patch is displayed in Fig. 5 (h). This conditioning helps the pCCGAN in optimal translation and restoration of the input CT image equivalent to the target MRI image as it utilizes the neighborhood pixel information also, both the high and low-frequency information is intact compared to the target modality. Fig. 5 (i) shows the result obtained by conditioning on patch size of $32\times32$ pixels. The low-frequency regions are not enhanced in this case, however, the image translation is comparable to the target image. The output obtained from conditioning on patch size of $64\times64$ is shown in Fig. 5 (j). In this case, the image contrast is increased and the high-frequency region is enhanced by a large amount. An ablation analysis with and without DiL is presented in Fig. \ref{3.5} (a) to (c). The restored MRI with DiL shown in Fig. \ref{3.5} (c) on the given MRI (Fig. \ref{3.5} (a)) successfully reduces the noise content and artifacts present in the low and frequency portion of the images compared to the restored version without DiL (Fig. \ref{3.5} (b)).
\section{Conclusion}
We have presented a novel patch-level cyclic conditional generative adversarial network embedded with adaptive dictionary learning; that successfully performs unpaired cyclic CT-MRI image translation-cum-restoration. The idea is based on learning underlying parameters for a stable cyclic transformation of random fields. We found that concatenating alternate patches (of appropriate size) from input and target scans helps learning hidden patterns of target image modality while preserving information features of input modality when back-translated. The model has not been tested for diagnostic power on the target images. This network finds application as first-analysis as well as practically useful in simulating target images when the required scanning machine is unavailable and/or unaffordable. In future, a unified neural network can be constructed for the unpaired translation of volumetric data (3D) among more than two medical imaging modalities.  
\bibliographystyle{IEEEtran}
\bibliography{refs}

\begin{thebibliography}{10}
\providecommand{\url}[1]{#1}
\csname url@samestyle\endcsname
\providecommand{\newblock}{\relax}
\providecommand{\bibinfo}[2]{#2}
\providecommand{\BIBentrySTDinterwordspacing}{\spaceskip=0pt\relax}
\providecommand{\BIBentryALTinterwordstretchfactor}{4}
\providecommand{\BIBentryALTinterwordspacing}{\spaceskip=\fontdimen2\font plus
\BIBentryALTinterwordstretchfactor\fontdimen3\font minus \fontdimen4\font\relax}
\providecommand{\BIBforeignlanguage}[2]{{%
\expandafter\ifx\csname l@#1\endcsname\relax
\typeout{** WARNING: IEEEtran.bst: No hyphenation pattern has been}%
\typeout{** loaded for the language `#1'. Using the pattern for}%
\typeout{** the default language instead.}%
\else
\language=\csname l@#1\endcsname
\fi
#2}}
\providecommand{\BIBdecl}{\relax}
\BIBdecl

\bibitem{vapnik2019complete}
V.~N. Vapnik, ``Complete statistical theory of learning,'' \emph{Automation and Remote Control}, vol.~80, no.~11, pp. 1949--1975, 2019.

\bibitem{rai2021augmented}
{Swati Rai, Jignesh S. Bhatt, and Sarat Kumar Patra}, ``Augmented noise learning framework for enhancing medical image denoising,'' \emph{IEEE Access}, vol.~9, pp. 117\,153--117\,168, 2021.

\bibitem{deshpande2019bayesian}
V.~S. Deshpande and J.~S. Bhatt, ``Bayesian deep learning for deformable medical image registration,'' in \emph{Pattern Recognition and Machine Intelligence: 8th International Conference, PReMI 2019, Tezpur, India, December 17-20, 2019, Proceedings, Part II}.\hskip 1em plus 0.5em minus 0.4em\relax Springer, 2019, pp. 41--49.

\bibitem{khalid2014single}
M.~U. Khalid and A.-K. Seghouane, ``A single svd sparse dictionary learning algorithm for fmri data analysis,'' in \emph{2014 IEEE Workshop on Statistical Signal Processing (SSP)}.\hskip 1em plus 0.5em minus 0.4em\relax IEEE, 2014, pp. 65--68.

\bibitem{naderahmadian2018generalized}
Y.~Naderahmadian and S.~Beheshti, ``Generalized adaptive weighted recursive least squares dictionary learning for retinal vessel inpainting,'' in \emph{2018 IEEE Statistical Signal Processing Workshop (SSP)}.\hskip 1em plus 0.5em minus 0.4em\relax IEEE, 2018, pp. 40--44.

\bibitem{rai2022accessible}
{Swati Rai, Jignesh S. Bhatt, and Sarat Kumar Patra}, ``Accessible, affordable and low-risk lungs health monitoring in covid-19: Deep cascade reconstruction from degraded lr-uldct,'' in \emph{2022 IEEE 19th International Symposium on Biomedical Imaging (ISBI)}.\hskip 1em plus 0.5em minus 0.4em\relax IEEE, 2022, pp. 1--5.

\bibitem{cherukuri2019deep}
V.~Cherukuri, T.~Guo, S.~J. Schiff, and V.~Monga, ``Deep mr brain image super-resolution using spatio-structural priors,'' \emph{IEEE Transactions on Image Processing}, vol.~29, pp. 1368--1383, 2019.

\bibitem{pratondo2015robust}
A.~Pratondo, C.-K. Chui, and S.-H. Ong, ``Robust edge-stop functions for edge-based active contour models in medical image segmentation,'' \emph{IEEE Signal Processing Letters}, vol.~23, no.~2, pp. 222--226, 2015.

\bibitem{armanious2019unsupervised}
K.~Armanious, C.~Jiang, S.~Abdulatif, T.~K{\"u}stner, S.~Gatidis, and B.~Yang, ``Unsupervised medical image translation using cycle-medgan,'' in \emph{2019 27th European Signal Processing Conference (EUSIPCO)}.\hskip 1em plus 0.5em minus 0.4em\relax IEEE, 2019, pp. 1--5.

\bibitem{zhu2020cross}
Y.~Zhu, Y.~Tang, Y.~Tang, D.~C. Elton, S.~Lee, P.~J. Pickhardt, and R.~M. Summers, ``Cross-domain medical image translation by shared latent gaussian mixture model,'' in \emph{International conference on medical image computing and computer-assisted intervention}.\hskip 1em plus 0.5em minus 0.4em\relax Springer, 2020, pp. 379--389.

\bibitem{liu2021ct}
Y.~Liu, A.~Chen, H.~Shi, S.~Huang, W.~Zheng, Z.~Liu, Q.~Zhang, and X.~Yang, ``Ct synthesis from mri using multi-cycle gan for head-and-neck radiation therapy,'' \emph{Computerized Medical Imaging and Graphics}, vol.~91, p. 101953, 2021.

\bibitem{NIPS2014_5ca3e9b1}
I.~Goodfellow, J.~Pouget-Abadie, M.~Mirza, B.~Xu, D.~Warde-Farley, S.~Ozair, A.~Courville, and Y.~Bengio, ``Generative adversarial nets,'' in \emph{Advances in Neural Information Processing Systems}, Z.~Ghahramani, M.~Welling, C.~Cortes, N.~Lawrence, and K.~Weinberger, Eds., vol.~27.\hskip 1em plus 0.5em minus 0.4em\relax Curran Associates, Inc., 2014.

\bibitem{xu2021domain}
W.~Xu and G.~Wang, ``A domain gap aware generative adversarial network for multi-domain image translation,'' \emph{IEEE Transactions on Image Processing}, vol.~31, pp. 72--84, 2021.

\bibitem{rai2021unsupervised}
{Swati Rai, Jignesh S. Bhatt, and Sarat Kumar Patra}, ``An unsupervised deep learning framework for medical image denoising,'' \emph{arXiv preprint arXiv:2103.06575}, 2021.

\bibitem{thummerer_adrian_2023_7260705}
\BIBentryALTinterwordspacing
A.~Thummerer, E.~van~der Bijl, and M.~Maspero, \emph{{SynthRAD2023 Grand Challenge dataset: synthetizing computed tomography for radiotherapy}}, Apr. 2023. [Online]. Available: \url{https://doi.org/10.5281/zenodo.7260705}
\BIBentrySTDinterwordspacing

\bibitem{CTMRI}
\BIBentryALTinterwordspacing
K.~Clark, B.~Vendt, K.~Smith, J.~Freymann, J.~Kirby, P.~Koppel, S.~Moore, S.~Phillips, D.~Maffitt, M.~Pringle, L.~Tarbox, and F.~Prior, \emph{The Cancer Imaging Archive (TCIA): Maintaining and Operating a Public Information Repository}, 2013. [Online]. Available: \url{https://doi.org/10.1007/s10278-013-9622-7}
\BIBentrySTDinterwordspacing

\end{thebibliography}
\end{document}